\documentclass[12pt]{iopart}

\usepackage{iopams}  
\usepackage{epsfig}
\begin{document}

\title{Charged hadron $R_{AA}$ as a function of $p_T$ at LHC}

\author{T.~Renk and K.~J.~Eskola}

\address{Department of Physics, P.O. Box 35 FI-40014 University of Jyv\"askyl\"a, Finland and Helsinki Institute of Physics, P.O. Box 64 FI-00014, University of Helsinki, Finland}
\ead{trenk@phys.jyu.fi}

\begin{abstract}
We compute the nuclear suppression factor $R_{AA}$ for charged hadrons within a radiative energy loss picture using a hydrodynamical evolution to describe the soft medium inducing energy loss. A minijet + saturation picture provides initial conditions for LHC energies and leading order perturbative QCD (LO pQCD) is used to compute the parton spectrum before distortion by energy loss.
\end{abstract}


We calculate the suppression of hard hadrons induced by the presence of a soft medium produced in central Pb-Pb collisions at $\sqrt{s} = 5.5$ TeV at the LHC. Note that this prediction depends on knowledge of the medium. In the present calculation, the medium evolution is likewise predicted and has to be confirmed before the suppression can be tested. Note further that the calculation is only valid where hadron production is dominated by fragmentation and that it cannot be generalized to the suppression of jets since the requirement of observing a hard hadron leads to showers in which the momentum flow is predominantly through a single parton. This is not so for jets in which the momentum flow is shared on average among several partons (which requires a different framework).

We describe the soft medium evolution by the boost-invariant 
hydrodynamical model discussed in \cite{Hydro} where the initial conditions for LHC are computed 
from perturbative QCD+saturation 
\cite{EKRT}. Our calculation for the propagation of partons through the medium 
follows the BDMPS formalism for radiative energy loss 
 using quenching weights \cite{QuenchingWeights}.
Details of the implementation can be found in \cite{Correlations}. 

The probability density $P(x_0, y_0)$ for finding a hard vertex at the 
transverse position ${\bf r_0} = (x_0,y_0)$ and impact 
parameter ${\bf b}$ is given by the normalized product of the nuclear profile functions.
We compute the energy loss 
probability $P(\Delta E)_{path}$ for any given path from a vertex through the medium by 
evaluating the line integrals
\begin{equation}
\label{E-omega}
\omega_c({\bf r_0}, \phi) = \int_0^\infty  d \xi \xi \hat{q}(\xi) \quad  {\rm and} \quad \langle\hat{q}L\rangle ({\bf r_0}, \phi) = \int_0^\infty d \xi \hat{q}(\xi).
\end{equation}
Along the path where we assume the relation
\begin{equation}
\label{E-qhat}
\hat{q}(\xi) = K \cdot 2 \cdot \epsilon^{3/4}(\xi) (\cosh \rho - \sinh \rho \cos\alpha)
\end{equation}
between the local transport coefficient $\hat{q}(\xi)$, the energy density $\epsilon$ and the local flow rapidity $\rho$ as given in the hydrodynamical model. The angle $\alpha$ is between flow and parton trajectory. 
We view  the constant $K$ as a tool to account for the uncertainty in the selection of $\alpha_s$ and possible non-perturbative effects increasing the quenching power of the medium (see \cite{Correlations}) and adjust it such that pionic $R_{AA}$ for central Au-Au collisions  at RHIC is described. The result for LHC is then an extrapolation with $K$ fixed.

Using the numerical results of \cite{QuenchingWeights}, we obtain $P(\Delta E; \omega_c, R)_{path}$ 
for $\omega_c$ and $R=2\omega_c^2/\langle\hat{q}L\rangle$.
From this distribution given a single path, we can define the averaged energy loss probability distribution $P(\Delta E)\rangle_{T_{AA}}$ by averaging over all possible paths, weighted with the probability density $P(x_0, y_0)$ for finding a hard vertex in the transverse plane.

We consider all partons as absorbed whose energy loss is formally larger than their initial energy.
The momentum spectrum of produced partons is calculated in LO pQCD. The medium-modified perturbative production of hadrons is obtained from the convolution
\begin{equation}
d\sigma_{med}^{AA\rightarrow h+X}  = \sum_f d\sigma_{vac}^{AA \rightarrow f +X} \otimes \langle P(\Delta E)\rangle_{T_{AA}} \otimes
D_{f \rightarrow h}^{vac}(z, \mu_F^2)
\end{equation} 
with $D_{f \rightarrow h}^{vac}(z, \mu_F^2)$ the fragmentation function. From this we compute the nuclear modification factor $R_{AA}$  as
\begin{equation}
R_{AA}(p_T,y) = \frac{dN^h_{AA}/dp_Tdy }{T_{AA}({\bf b}) d\sigma^{pp}/dp_Tdy}.
\end{equation}

\begin{figure}[htb]
\begin{center}
\epsfig{file=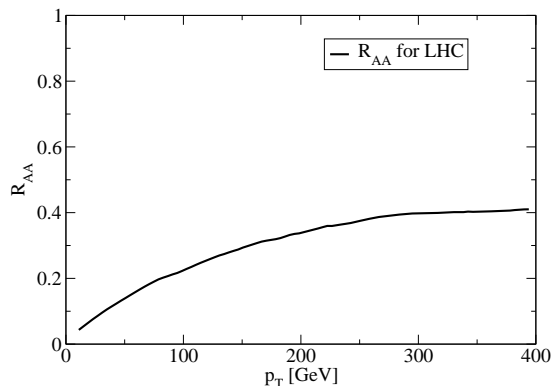, width=7.2cm} \vspace*{-0.8cm}
\end{center}
\caption{\label{F1}Expectation for the $p_T$ dependence of the nuclear suppression factor $R_{AA}$ for charged hadrons in central Pb-Pb collisions at midrapidity at the LHC.}
\end{figure}

Fig.~\ref{F1} shows the expected behaviour of $R_{AA}$ with hadronic transverse momentum $p_T$ at midrapidity. On quite general grounds, we expect a rise of $R_{AA}$ with $p_T$ for any energy loss model in which the energy loss probability does not strongly depend on the initial parton energy as more of the shift in energy becomes accessible (see \cite{Correlations}). The detailed form of the rise is then sensitive to the form of $P(\Delta E)\rangle_{T_{AA}}$.

\end{document}